\documentclass[twocolumn,showpacs,preprintnumbers,amsmath,amssymb,superscriptaddress]{revtex4}

\usepackage{graphicx}
\usepackage{dcolumn}
\usepackage{bm}
\newcommand{\nc}{\newcommand}
\nc{\be}{\begin{equation}}
\nc{\ee}{\end{equation}}
\nc{\bea}{\begin{eqnarray}}
\nc{\eea}{\end{eqnarray}}
\nc{\bean}{\begin{eqnarray*}}
\nc{\eean}{\end{eqnarray*}}
\nc{\mb}{\mbox}
\nc{\rnc}{\renewcommand}
\nc{\vk}{\mb{\bf k}}
\nc{\vp}{\mb{\bf p}}
\nc{\vn}{\mb{\bf n}}
\nc{\vq}{\mb{\bf q}}
\nc{\rr}{\mb{\bf r}}
\nc{\xx}{{\mb{\bf x}}}
\nc{\vz}{\hat {\mb{\bf z}}}
\nc{\vjj}{\mb{\boldmath$j$}}
\nc{\vg}{\mb{\boldmath$g$}}
\nc{\x}{\mb{\boldmath$x$}}
\nc{\A}{\mb{\boldmath$A$}}
\nc{\va}{\mb{\boldmath$a$}}
\nc{\vs}{\mb{\boldmath$\sigma$}}
\nc{\vpi}{\mb{\boldmath$\pi$}}
\nc{\nab}{\nabla}
\nc{\X}{\sf x}

\begin{document}


\title{Edge spin accumulation in semiconductor two-dimensional hole gases}
\author{K. Nomura}
\affiliation{Department of Physics, University of Texas at Austin,
Austin TX 78712-1081, USA}
\author{J. Wunderlich}
\affiliation{Hitachi Cambridge Laboratory, Cambridge CB3 0HE, UK}
\author{Jairo Sinova}
\affiliation{Department of Physics, Texas A\&M University, College
Station, TX 77843-4242, USA}
\author{B. Kaestner}
\affiliation{Hitachi Cambridge Laboratory, Cambridge CB3 0HE, UK}
\affiliation{National Physical Laboratory, Teddington T11 0LW, UK}
\author{A.H. MacDonald}
\affiliation{Department of Physics, University of Texas at Austin,
Austin TX 78712-1081, USA}
\author{T. Jungwirth}
\affiliation{Institute of Physics ASCR, Cukrovarnick\'a 10, 162 53
Praha 6, Czech Republic}
\affiliation{School of Physics and Astronomy, University of Nottingham,
University Park, Nottingham NG7 2RD, UK} 

\date{\today}

\begin{abstract}
{\bf The controlled generation of localized spin densities is a key enabler of 
semiconductor spintronics 
In this work, we study spin Hall effect  induced edge spin accumulation in a two-dimensional 
hole gas  with strong spin orbit  interactions. We argue that it is an intrinsic property, 
in the sense that it is independent of the strength of disorder scattering. 
We show numerically that the spin polarization near the edge induced by this mechanism can be
large, 
and that it becomes larger and more strongly localized as the spin-orbit coupling strength increases, and is independent
of the width of the conducting strip once this exceeds the elastic scattering mean-free-path.
Our experiments in two-dimensional hole gas microdevices 
confirm this remarkable spin Hall effect phenomenology. 
Achieving comparable levels of spin polarization by external magnetic fields would require
laboratory equipment whose physical dimensions and operating electrical currents
are million times larger than those of our spin Hall effect devices.
}
\end{abstract}


\maketitle
\section{Introduction}
Finite spin densities in semiconductors have traditionally \cite{Zutic:2004_a}
been generated by external magnetic fields, by circularly polarized light sources, 
or by spin injection from ferromagnets.  Recently there has been considerable interest \cite{Day:2005_a}
in an alternate strategy in which edge spin densities are generated electrically via the spin Hall effect (SHE)
\cite{Kato:2004_d,Wunderlich:2004_a},
i.e., in a planar device by the current of spins  oriented perpendicular to the plane that is generated 
by and flows perpendicular to an electric field.  
The SHE has traditionally been thought of as a consequence of  
spin-dependent chirality in impurity scattering that occurs 
in systems with spin-orbit (SO) coupling \cite{Dyakonov:1971_b,Hirsch:1999_a}. 
Recently it has been recognized that the SHE also has an intrinsic contribution
due to SO coupling in a perfect crystal \cite{Murakami:2003_a,Sinova:2004_a}.
The theoretical works \cite{Hirsch:1999_a,Murakami:2003_a,Sinova:2004_a} 
and the subsequent experimental discovery
\cite{Kato:2004_d,Wunderlich:2004_a} of the SHE
in both strong and weak SO interaction regimes have generated substantial interest in both 
the origins of this unfamiliar transport phenomenon and its potential for low power consumption
electronic devices \cite{Murakami:2003_a,Kato:2004_d}.
The role of disorder in the spin-Hall conductivity is more subtle than in more familiar
transport coefficients. Early theoretical work reflected a variety of strongly polarized 
views, all the way from a picture with large dissipationless
transverse spin currents \cite{Murakami:2003_a,Sinova:2004_a} in which disorder
plays no role, to demonstrations that the effect is completely eliminated by infinitesimally weak disorder 
\cite{Inoue:2004_a,Mishchenko:2004_a,Raimondi:2004_a,Chalaev:2004_a}. 
The theoretical picture is now becoming more clear, if more complex. 
It is now established that many of the theoretical controversies have arisen 
because what is seemingly the simplest possible model system, a two-dimensional
electron gas with Rashba SO coupling \cite{Rashba:1960_a},
turns out to have anomalous properties.  Although the SHE is absent 
in this model, it is present and robust against disorder
\cite{Murakami:2005_a,Mal'shukov:2004_a,Bernevig:2005_a,Nomura:2005_a} in generic systems and 
in all other models that have been specifically studied.  

Another challenging issue in SHE theory concerns the definition of the spin current in SO coupled systems; since 
spin is not conserved the merit of a particular definition has to be decided on the basis of its utility 
for the evaluation of observable properties.  The fact that different definitions have been proposed which yield
spin Hall currents that differ even by a sign for some models emphasizes the importance of this 
point \cite{Rashba:2003_a,Culcer:2004_a,Zhang:2005_b}.  The most important observable associated with the 
SHE is the edge spin accumulations to which it gives rise; indeed experiments that discovered the SHE
\cite{Kato:2004_d,Wunderlich:2004_a} did so by measuring edge spin accumulations.  These measurements 
have partly steered the focus of theoretical studies from the experimentally ambiguous spin Hall currents to 
 spin Hall accumulations \cite{Ma:2004_a,Nikolic:2004_b,Wang:2005_b,Onoda:2005_b}. 
The opportunity to directly compare
theoretical and experimental values of a SHE related physical quantity has been appealing, yet physicists have 
entered this territory armed with no analytical theory that is able to predict the sign, magnitude, and spatial 
extent of the SHE spin accumulation, at least in the most interesting case of strong SO interactions.
Particularly intriguing in this context is the dual role of SO coupling which
generates the spin-polarization of carriers while, at the same time, causing spin decoherence
upon scattering off channel edges or off impurities. 
In the present paper we compare a numerical study and experimental results of the SHE spin accumulation in 
two-dimensional hole gases (2DHGs). 
The systems we study are in the diffusive transport regime in which the sample dimensions are 
large compared to scattering mean-free-path.

\section{Theory}
First we discuss our theoretical results. The calculations are based on the following 
model Hamiltonian \cite{Winkler:2000_a,Schliemann:2004_a} of strongly confined 2DHG at semiconductor heterojunction with structural inversion asymmetry:
\bea
H_R&=&
\left(
\begin{array}{rr}
\hbar^2 k^2/2m & \ \ \ \ i\lambda k_-^{3}\ \   \\
 \ \ -i\lambda k_+^{3}\ \  &  \hbar^2 k^2/2m
\end{array}
\right)\; ,
\label{H_R}
\eea
where $m$ is the carrier effective mass and the momenta $k=(k_x^2+k_y^2)^{1/2}$ and $k_{\pm}=k_x\pm ik_y$.
The 2DHG strip is defined by hard wall barriers separated by a distance $L_x=L $ in the transverse $x$-direction and
by a longitudinal length $L_y=L $, and periodic boundary condition in the
$y$-direction.
Disorder is modeled assuming uncorrelated, short-range spin-independent scatterers, i.e.,  
$
V(\rr)=\sum_{I=1}^{N_i} V\delta(\rr-{\bf R}_I)
$.
This "white-noise" disorder potential satisfies 
$\langle V(\vq) V(\vq')^* \rangle=N_iV^2\delta(\vq-\vq'),$
 where $V(\vq)$ is the Fourier component of $V(\rr)$ and $N_i=n_iL^2$ is the number of impurities; $n_i\approx n$
 where $n$ is the 2DHG density.  The strength of disorder can be characterized
 by the corresponding Born approximation scattering rate, $\hbar/\tau=2\pi
 n_iV^2\rho_F$, where $\rho_F$ is the 2D density of states in the
 absence of SO coupling.
In the following we relate all energies to the
Fermi energy, $E_F$, of the 2DHG in the absence of SO coupling 
and disorder, and all lengths to the corresponding inverse Fermi wavevector $k_F^{-1}$. In our studies,
$E_F$  is always the largest relevant energy scale. 
\begin{figure}[t!]
\begin{center}
\includegraphics[width=0.48\textwidth]{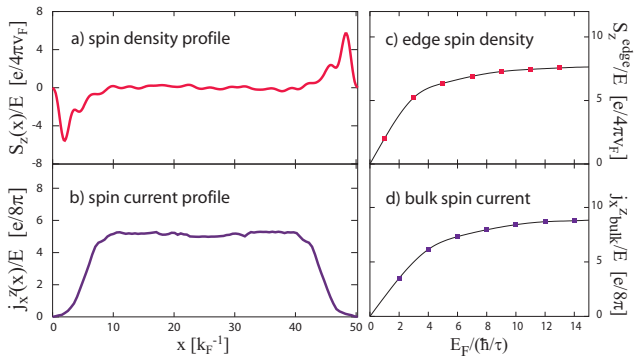}
\vspace{-0.5cm}
\caption{
Spin Hall effect in a 2DHG. 
Spatial profile of the $z$-component of spin (a) and of the spin current (b) across the channel width
($x$-direction). Disorder dependence of the accumulated edge spin density (c) and of bulk spin current (d).
}
\label{fig1}
\end{center}
\end{figure}

The SHE spin-accumulation is
generated near the edges of the system by a charge current in the $y$-direction.
Linear response theory predicts an induced spin density given by the Kubo formula,
\bea
 S_z(x)&=&-i\hbar E_y\int_0^{L_y}\frac{dy}{L_y}\sum_{\alpha,\alpha'}\frac{f(E_\alpha)-f(E_{\alpha'})}{E_\alpha-E_{\alpha'}}\nonumber \\
&& \frac{\langle \alpha|s^z(x,y)|\alpha'\rangle \langle \alpha'|j_y|\alpha\rangle}{E_\alpha-E_{\alpha'}+i\eta}\; ,
\label{eq:kuboI} 
\eea
where $s^z(x,y)$ is the $z$-component of the local spin density operator (we assume spin $3\hbar/2$ for the heavy-holes), $j_y=e\partial H/\partial \hbar k_y$ is the longitudinal electrical current operator for holes,
and $E_\alpha$ and $|\alpha\rangle$ are the eigenenergies and eigenstates of the full Hamiltonian $H$ which includes the SO coupling term 
(\ref{H_R}), the disorder potential, and the hard-wall confining potential.
 $\eta$ must be chosen to be 
small compared to all intensive energy scales, such as
$E_F$, $\hbar/\tau$, and the characteristic
SO coupling strength $\Delta_{so}=2\lambda k_F^{3}$, but larger than the finite system level spacing.
A detailed discussion of subtleties related to the proper choice of $\eta$ can be found in Ref.~\cite{Nomura:2005_a}.

Typical results of the calculated spatial profile of spin-density induced by a longitudinal electric
field $E_y$ are shown in  Fig.~\ref{fig1}(a).
For comparison we show in  Figs.~\ref{fig1}(b) the 
transverse spin current, $j_x^z$,  also obtained using linear response theory \cite{Sinova:2004_a,Schliemann:2004_a,Nomura:2005_a}.
In both panels, $\Delta_{so}/E_F=0.4$, 
$(\hbar/\tau)/E_F=0.3$, and the diffusive regime is obtained by assuming  system size $L= 50[k_F^{-1}]$
which is  
larger than the mean free path,
$l=6[k_F^{-1}]$.
The $z$-component of the induced spin polarization has maxima near the strip edges, while the spin
current has a maximum in the middle of the sample and decays toward the edges.
We note that the peak value of the spin current divided by the applied electric field
is comparable to the spin Hall conductivity calculated in a previous study
with periodic boundary conditions rather than hard walls in the $x$-direction \cite{Nomura:2005_a}.

In Figs.~\ref{fig1}(c) and (d) we illustrate the disorder dependence of the peak value
of the edge spin density, $S^z_{edge}$,  
and of the transverse
spin current in the middle of the strip, $j^z_{x,bulk}$.
With increasing disorder strength, both $S^z_{edge}$ and $j^z_{x,bulk}$
decrease gradually and vanish in the strong
scattering limit, $\hbar/\tau\gg\Delta_{so}$. 
The striking similarity of the dependences on disorder strength of the two quantities
led us to a numerical observation that, 
\bea
 v_F \; S^z_{\rm edge}\sim j^{z}_{\rm bulk},
 \label{Svsj}
\label{spin1}
\eea
where $v_F$ is the hole Fermi velocity. Introducing the characteristic SO coupling length $L_{so}=v_F\pi\hbar/\Delta_{so}$ and
time  $t_{so}=\hbar\pi/\Delta_{so}$, we can rewrite Eq.~(\ref{Svsj}) as
\bea
 L_{so}S^z_{\rm edge}\sim t_{so}j^{z}_{\rm bulk}.
\label{spin2}
\eea
Here we emphasize that both $L_{so}$ and
$t_{so}$ have an intrinsic, disorder independent origin. Eq.~(\ref{spin2}) suggests 
that the spin polarization accumulated in the region of size $L_{so}$, $L_{so}S^z_{edge}$, is proportional to
the amount of incoming spin current from the middle of the sample in a
time scale $t_{so}$.

\begin{figure}[!t]
\begin{center}
\includegraphics[width=0.48\textwidth]{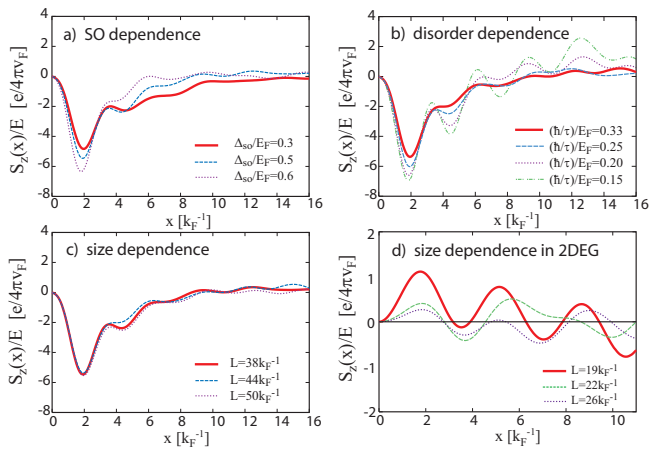}
\vspace{-0.5cm}
\caption{
Spatial profile of the $z$-component of accumulated spin density as a function of 
  spin-orbit coupling for $(\hbar/\tau)/E_F=0.3$ and $L=44k_F^{-1}$ (a), the disorder strength 
  for $\Delta_{so}/E_F=0.4$ and $L=44k_F^{-1}$ (b), and of the system size for 
  $(\hbar/\tau)/E_F=0.3$ and $\Delta_{so}/E_F=0.4$ (c). System size
 dependence in a 2DEG for $(\hbar/\tau)/E_F=0.15$ and
 $\Delta_{so}/E_F=0.4$ (d).}
\label{fig2}
\end{center}
\end{figure}

To further examine the validity of this {\em ansatz}~(\ref{spin2})
we plot in Figs.~\ref{fig2}(a) and (b) the dependence of the spin polarization profile on
$\Delta_{so}$ and on $\hbar/\tau$.
As seen in Fig.~\ref{fig2}(a), the spin density builds up more gradually from the center
towards the edge of the strip for weaker SO coupling.
Fig.~\ref{fig2}(b) demonstrates explicitly that the envelope of 
the spin polarization curve changes only weakly with disorder.  
Calculations presented in Fig.~\ref{fig2}(c) further demonstrate that the edge spin accumulation
properties do not change once the system size exceeds the SO length $L_{so}$.

Our findings on the disorder dependence of spin-accumulation differ from 
expectations based on analogies to other regimes.  In the case of strong
disorder 
the accumulation time scale is the spin-relaxation time \cite{Dyakonov:1971_b} $\tau_{s} \sim \Delta_{so}^{-2} \tau^{-1}$ 
while the accumulation length scale is the spin-diffusion length 
$l_s=\sqrt{D\tau_s}$ where $D$ is the diffusion coefficient.
For a given spin Hall current, analogies with this regime
predict accumulation on the spin diffusion length $l_s$\cite{Murakami:2003_a,Culcer:2004_a,Zhang:2005_b} or the spin-precession length scale
$L_{so}$\cite{Mal'shukov:2005_a}, which corresponds with $l_s$ in the
dirty limit\cite{Dyakonov:1971_b}, and predict that the
magnitude is proportional to $\tau_s\propto \tau^{-1}$. The applicability of this analogy in the strong SO 
scattering limit is 
not established theoretically, however, and our numerical results suggest that it is doubtful. 
Consistency is not achieved by simply replacing the weak SO interaction 
expression for the spin relaxation time by the strong SO interaction 
result that $\tau_{s}\sim\tau$, which would imply that the spin accumulation  
is proportional to $\tau$\cite{Murakami:2003_a,Culcer:2004_a,Zhang:2005_b}.  

\begin{figure}[t!]
\begin{center}
\includegraphics[width=0.48\textwidth]{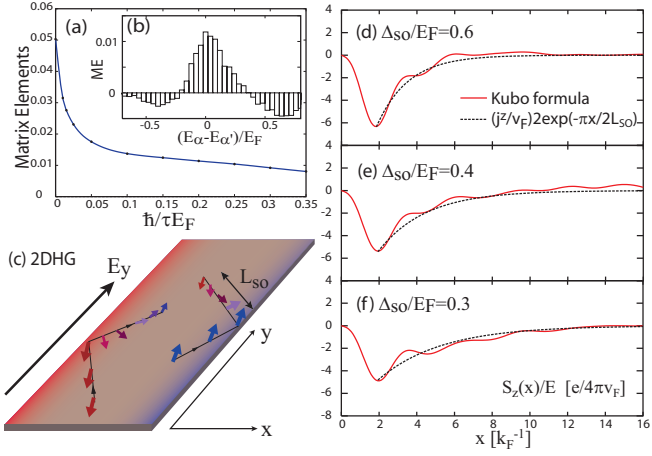}
\caption{
 Matrix elements that appear in Eq.~(\protect\ref{boo}) as a function of the scattering rate 
 at
 $E_{\alpha}-E_{\alpha^{\prime}}=0$ (a) and as a function of
the energy difference at $(\hbar/\tau)/E_F=0.14$ (b). $\Delta_{so}/E_F=0.3$ and 
$L=44k_F^{-1}$ in both panels.
(c) Shematic of the spin Hall accumulation in the strong SO coupling
 regime.
 Spatial profile of $z$-component of spin density for (d)
 $\Delta_{so}/E_F=0.6$, 0.4 (e), and 0.3 (f) at $(\hbar/\tau)/E_F=0.3$.
}
\label{fig3}
\end{center}
\end{figure}

We can understand why the Kubo formula predicts that the induced spin-density is 
weakly dependent on scattering rate 
in the diffusive limit by considering the matrix elements of the spin-density and 
current operators in Eq.~(\ref{eq:kuboI}).  
Because the paramagnetic system is even under time-reversal and the spin-density is odd, it 
follows that its static linear response originates from the odd dissipative term in the Kubo formula  
\begin{eqnarray}
 S_z(x)&=&-\pi \hbar E_y\int_0^{L_y}\frac{dy}{L_y}
\sum_{\alpha,\alpha'}\frac{\partial f(E_{\alpha})}{\partial E_\alpha }\nonumber \\
& & \langle \alpha|s^z(x,y)|\alpha'\rangle \langle \alpha'|j_y|\alpha\rangle 
\delta_{\eta}(E_\alpha-E_{\alpha'}),
\label{boo}
\end{eqnarray}
where $\delta_{\eta}(E) \equiv \frac{\eta}{\pi [ E^2 + \eta^2]}$
Typical properties of these 
matrix elements are illustrated in Fig.~\ref{fig3} where by "ME" we denote the above product of the spin and current
matrix elements.
As in all transport properties, current operator matrix elements in the absence of 
disorder play a key role.  For 2DHG strips, the component of momentum along the strips 
is a good quantum number whereas the transverse momentum is replaced by
discretestanding wave or one-dimensinal 
subband indices.  Both spin and charge operators 
have matrix elements that are diagonal in the 1D wavevector and in subband index.  It 
follows that the induced spin-density that we calculate is (at $T=0$) proportional to 
$\eta^{-1}$ in the absence of disorder, diverging when the $\eta \to 0$ limit is taken.
Unsurprisingly disorder, which limits the lifetime of these 1D band states, must be 
considered to obtain a finite result.  In the limit of the quasi 1D
limit or very weak disorder, the lifetime
broadening by ($\hbar / \tau$) of the band states is smaller than the subband separation,
and it follows from Eq.~(\ref{boo}) that the induced spin-density will be proportional to $\tau$ \cite{Governale:2002_a,Usaj:2004_a}.
This is the limit in which the 1D subbands contribute independently to the induced spin density and
can be treated as separate bands in a Boltzmann transport theory.

The 1D band limit applies only when $\hbar/\tau$ is smaller than the subband splitting,
i.e., when $\hbar/\tau < \hbar v_F / L_x$ and therefore when the mean free path 
$l = v_F \tau $ is larger than $L_x$.  As emphasized above, our calculations were 
performed in the regime in which $l$ is small compared to the strip width. 
As illustrated in Fig.~\ref{fig3}, intersubband transitions contribute to the induced spin-density;
electric field induced intersubband coherence cannot be neglected.  Although the contribution from each transition 
is proportional to $\tau$, the number of transitions that are incoherently mixed and contribute 
is proportional to $\tau^{-1}$, leading to our finding of weak $\tau$ dependence.

We found that in the diffusive region, in which the system size is larger
than the mean-free path, the matrix element products that appears in
Eq.~(\ref{boo}) at $E_{\alpha}-E_{\alpha'} \to 0$ changes slowly with 
disorder.  Exact disorder eigenstates at nearby energies mix 1D band states with disorder 
free energies that vary over the range $\hbar / \tau$ and in the regime of interest 
strongly mix states belonging to different subbands.
Eq.~(\ref{boo}) shows that the induced spin-density is proportional to 
the quantity plotted in the main panel of Fig. ~\ref{fig3}(a), which varies
rapidly with scattering rate when the mean free
path exceeds the system size, i.e. low values of $(\hbar/\tau)/E_F$, and
 slowly at larger scattering rates as expected from the 
above argument. Fig.~\ref{fig3}(b) shows the eigenenergy difference dependence of the 
matrix element product which remains large over the energy interval $\Delta_{so}$.

The above detailed analysis suggests the following physical picutre of
the spin Hall accumulation in the strong SO interaction regime.
Here we consider a reflection process from a state with a momentum
$\vk'$ to another state $\vk$ at the left boundary as shown in the
Fig.3. In the following we assume the width of stripe $L$ is infinitely
large, and focus on the major band. Note that the SO
Hamiltonian has a form 
$
H_{so}=-{\bf \Delta}(\vk)\cdot \vs/2
$
where
the effective Zeeman field is given as ${\bf
\Delta}(\vk)=2\lambda k^{\alpha}(-\sin(\alpha\phi),\cos(\alpha\phi),0)$ 
with $\alpha=3$ for 2DHGs and $\alpha=1$ for 2DEGs and $\phi=\tan^{-1}(k_y/k_x)$.
Thus in a steady state the spins are parallel to the direction of this
effective field ${\hat{\bf {\Delta}}}={\bf {\Delta}}/|{\bf {\Delta}}|$.
The dynamics of the direction of spins ${\bf m}={\bf S}/S$ is described
by the equation of motion
\bea
 \frac{\partial}{\partial t}{\bf m}(\vk,t)={\bf
 m}(\vk,t)\times
\frac{\bf \Delta(\vk)}{\hbar}
+\left(
\frac{e{\bf E}}{\hbar}
\cdot
\nabla_{\vk}
\right)
{\bf m}(\vk,t)\nonumber \\
\label{eqm}
\eea
in the clean limit.
The spin precesses around ${\bf \Delta}(\vk')$ until the reflection and
changes its direction of the precession axis to ${\bf \Delta}(\vk)$.
In the presence of the electric field $z$-component spins are
generated as discussed in the original paper of the intrinsic spin Hall
effect\cite{Sinova:2004_a}.
Assuming ${\bf m}(\vk,t)\simeq {\hat{\bf {\Delta}}}(\vk)+\delta{\bf
m}(\vk,t)$, where $\delta{\bf m}$ is linear in $eE$, we obtain 
$({\partial}/{\partial t})^2\delta m_z(\vk,t)=-|{\bf {\Delta}}/\hbar|^2(\delta
m_z(\vk,t)-m_z^{0}(\vk,{\bf E}))$ from Eq.(\ref{eqm}). Here note that the static component
\bea
m_z^{0}(\vk,{\bf E})=\alpha e{\bf E}
\cdot[\vz\times\hat{\vk}]/k|{\bf \Delta}(\vk)|
\eea
 causes the spin current density
$\vjj^z=\frac{1}{L^2}\sum_{\vk,M=\pm}f(E_{\vk,M})\left(M m^0_z(\vk,{\bf
E})\frac{s\vk}{m}\right)=\sigma_{sH}[{\bf E}\times\vz]$
in the bulk\cite{Sinova:2004_a}, where $s=3/2$ for 2DHGs.
Note that the initial
state with $\vk'$ 
has a negative z-component of spin 
$m_z^0=-\alpha eE|\hat{k}_x'|/ k'|{\bf {\Delta}}(\vk')|$, 
while the right moving
reflected state with $\vk$ has a positive 
$m_z^0=\alpha eE|\hat{k}_x|/k|{\bf {\Delta}}(\vk)|$. 
A typical amount of $z$-component
of spin for all left moving states ($k'_x<0$) is
$\frac{s}{L^2}\sum_{k'_x\le0,k'_y,M} 
f(E_{\vk',M}) M m^0_z(\vk',{\bf E})\simeq -s\alpha eE/4\pi v_F=
-|j^z|/v_F$, and the same amount with opposite sign $+|j^z|/v_F$ is for
all right moving
states ($k_x>0$). Therefore in the bulk of the sample
these components cancel each other. On the other hand, near the edge
this cancellation cannot be accomplished because it takes a finite time to
change the averaged z-component spin from $m_z^0(\vk',{\bf E})$ to
$m_z^0(\vk,{\bf E})$ by the reflection as indicated schematically in Fig.3(c).
The net spin density arises from this edge induced coherence between
different Bloch states and falls to zero on the length scale
$L_{so}2/\pi$ because of averaging over scattering angles as
\bea
  S_z(x)=-\frac{j^z_{\rm bulk}}{v_F}2\exp\left(-\frac{\pi}{2}\frac{x}{L_{so}}\right).
\eea
Here the factor $2/\pi$ comes from averaging of scattering angles.
As shown in Fig.3(d)-(f) this ansatz is in a very good agreement with our
numerical Kubo formula results.
Here we shifted the $x$ coordinate as $x\rightarrow x-x_{\rm edge}$,
where $x_{\rm edge}$ is the point of the peak of $S_z$.
These results show a qualitative correspondence between spin accumulation 
and the bulk SHE with the standard spin-current definition \cite{Sinova:2004_a}.  

We end this theoretical sections by noting that  qualitatively different behavior is expected for two-dimensional electron
systems with  Rashba SO coupling. 
Fig.~\ref{fig2}(d) shows oscillations of the $z$-component of the 2DEG spin density 
whose amplitude decreases with increasing system size. 
The lack of clear edge spin accumulation in this system we obtained numerically
is consistent with the suppressed bulk spin current of this model
\cite{Inoue:2004_a,Mishchenko:2004_a,Raimondi:2004_a,Chalaev:2004_a,Nomura:2005_a,Mal'shukov:2005_a}.
A clear edge spin accumulation could appear in the non-linear transport
regime with a strong electric field in a ballistic nanostructre.\cite{Nikolic:2004_b}

\begin{figure}[h]
\begin{center}
\vspace{-1cm}
\includegraphics[width=0.5\textwidth]{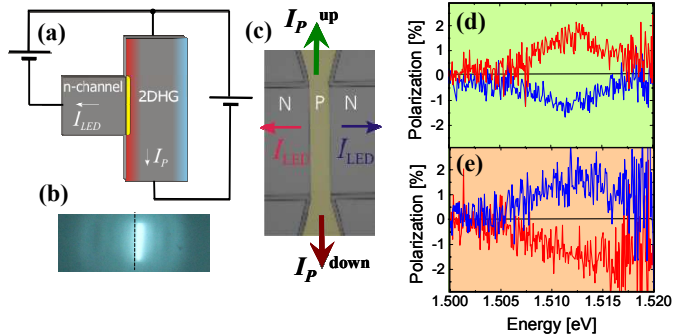}
\vspace{-1.5cm}
\caption{Schematic configuration of lateral p-n junction to detect spin accumulation (a): 
2DHG channel bordered by an n-type region which forms LED. Driving a current $I_{LED}$ through the LED results in electroluminescence indicated by the yellow line. Driving a current $I_p$ along the 2DHG channel induces
the SHE spin accumulations at the edges of the 2DHG. Light emission from the p-n junction recorded by a CCD camera (b). Electron microscope image of the microdevice with symmetrically placed LEDs at both
edges of the 2DHG channel (c). 
Emitted light polarization of recombined light in the p-n junction, (d) and (e), for the current flow indicated in (c).}
\label{fig5}
\end{center}
\end{figure}

\section{Experiment}
We now show that the above theoretical phenomenology of the SHE in strong SO coupling regime
is consistent with our experimental data in modulation doped, GaAs/AlGaAs 2DHG devices. To demonstrate
the independence of the SHE on the channel width we compare measurements in the 1.5~$\mu$m
channel \cite{Wunderlich:2004_a} and in a new device with a 10~$\mu$m wide channel, lithographically defined in
the same 2DHG system with $l\approx 50$~nm.
In Fig.~\ref{fig5}(a) we show schematically the experimental setup with the LED current, $I_{LED}$, driving the electroluminescence
at the edge of the hole strip,  and the 2DHG channel current, $I_p$, inducing the SHE edge spin accumulation. 
Digital camera image of the light emitted from the
$p-n$ junction is shown in Fig.~\ref{fig5}(b). Fig.~\ref{fig5}(c) is a scanning
electron microscope image of the  whole microdevice with symmetrically placed LEDs at both
edges of the 10~$\mu$m channel. 
Circular polarization of the spectral peak corresponding to recombination of electrons with the spin-polarized
2DHG states at the
edges of the channel are shown in Figs.~\ref{fig5}(d) and (e). For more details on our
co-planar LED devices and on the spectroscopical
analysis of the electroluminescence data see Refs.~\cite{Wunderlich:2004_a},\cite{Kaestner:2005_a}
 and citations therein.

Consistency of the measured data with the basic SHE phenomenology is demonstrated by the detected opposite sign of the $z$-component
of spin
accumulation at opposite edges (compare red and blue curves in Fig.~\ref{fig5}(d) or (e)) and by the
observed flipping of the sign of spin polarization 
at a given edge upon reversing $I_p$ (compare, e.g., red curves in Figs.~\ref{fig5}(d) and (e)).
The magnitude of the $I_p$ current in the 10~$\mu$m channel was adjusted so that the corresponding longitudinal 
electric field, $E_y\approx0.15$V$\mu$m$^{-1}$, was comparable to 
the field used in the experiment in the 1.5~$\mu$m
channel \cite{Wunderlich:2004_a}. Spin polarizations of order $\sim 1$\% observed in both devices confirm the expected
independence of the SHE signal on the width of the diffusive channel. 

We now compare experimental and calculated magnitudes of the edge spin polarization.
Numerical data in Fig.~\ref{fig2} imply an approximate general form of the dimensionless edge polarization,
\bea
\frac{S^z_{edge}}{\hbar n}=
g_s\frac{e}{4\pi v_F\hbar n}E_y,
\label{sz_num}
\eea
where $g_s$ is a numerical factor.
In the measured system the 2DHG density  $n=2\times 10^{12}$cm$^{-2}$, $v_F=10^5$ms$^{-1}$, $\Delta_{so}/E_F\approx0.4$,
and $(\hbar/\tau)/E_F \approx 0.1$. The system is therefore in the strong SO coupling regime and the corresponding 
numerical factor $g_s=7.4$. From Eq.~(\ref{sz_num}) we obtain that the expected polarization for the measured sample reaches $\sim 8$\% 
and the width of the accumulation area is of order $L_{so}\sim 10$nm.  The theoretical $\sim 8$\% polarization is consistent with the measured value
of $\sim 1$\% assuming an effective LED recombination width of the order $\sim 100$nm. This number cannot be precisely determined 
experimentally because of the resolution limit set up by the wavelength of the emitted light ($\sim 800$nm). However, comparisons between
experiments in the 1.5~$\mu$m and 10~$\mu$m 2DHG channels, analysis of the digital images of the active $p-n$ junction area, and simulations
of device 
$I-V$ characteristics \cite{Wunderlich:2004_a} confirm a sub-micron width of the recombination region near the $p-n$ junction.  

\section{Summary}
 In summary we have studied the edge spin accumulation due to the SHE  in a 2DHG 
 both theoretically and experimentally.
We found theoretically a clear correspondence between the
 edge spins and the bulk spin currents.
 The calculated spin-density is nearly independent of the disorder scattering rate in magnitude and is
 localized within a spin-precession length of the edge.  The magnitude increases with the strength of the 
 SO coupling. For the parameters used 
 in the experimental study, we predict an 8\% spin-polarization at the edge.  These predictions are 
 consistent with the experimental finding of $\sim 1$\% optical polarization averaged over a distance 
 $ \sim 10 L_{so}$.  The experiments also confirmed that opposite spin accumulations at the edges can be 
 be separated over large distances in diffusive conduction channels and that their magnitude is not affected by
 the channel width. 

 \section*{Acknowledment}
The authors would like to thank B.A. Nikolic and E. Rashba for useful discusstions.
 The work has been supported by
 the Japan Society for the Promotion of Science
Grant Agency and 
Academy of Sciences of the  Czech  Republic, EPSRC,
Welch Foundation, and Department of Energy.


\end{document}